\documentclass[a4paper,11pt,final]{article}

\usepackage{amsmath,amsfonts,amssymb,bm,mathptmx,graphicx}

\DeclareFontFamily{OT1}{pzc}{}
\DeclareFontShape{OT1}{pzc}{m}{it}{<-> s*[1.15] pzcmi7t}{}
\DeclareMathAlphabet{\mathcal}{OT1}{pzc}{m}{it}

\usepackage[small,bf]{caption}

\usepackage{hyperref}
\hypersetup{pdftitle={Exact eigenspectrum of the symmetric simple exclusion process on the complete, complete bipartite, and related graphs}, pdfauthor={J. Ricardo G. Mendonça}, pdfsubject={Exclusion process, statistical mechanics, Heisenberg model, complete graph}, pdfkeywords={Simple exclusion process, Heisenberg model, complete graph, Curie-Weiss model, SU(2) algebra, Clebsch-Gordan series}, unicode, extension=pdf, pdfnewwindow, pdftoolbar, pdfmenubar, pdffitwindow=false, pdfstartview={FitH}, colorlinks, linkcolor=blue, citecolor=blue, urlcolor=blue}
\def\leq{\leqslant}
\def\geq{\geqslant}
\def\I{\mbox{\rm i}}

\begin{document}

\pagestyle{empty}

\renewcommand{\thefootnote}{\fnsymbol{footnote}}

\begin{titlepage}

\begin{center}

{\Large \bf \makebox[0pt][c]{Exact eigenspectrum of the symmetric simple exclusion process} \\ on the complete, complete bipartite, and related graphs}

\vspace{6ex}

{\large {\bf J. Ricardo G. Mendon\c{c}a$^{a,b,}$}}\footnote{Email:
\href{mailto:jricardo@usp.br}{\nolinkurl{jricardo@usp.br}}.}

\vspace{1ex}

{\it $^{a}$Escola de Artes, Ci\^{e}ncias e Humanidades, Universidade de S\~{a}o Paulo \\ \makebox[0pt][c]{Avenida Arlindo B\'{e}ttio 1000, Ermelino Matarazzo -- 03828-000 S\~{a}o Paulo, SP, Brazil}}

\vspace{1ex}

{\it \makebox[0pt][c]{$^{b}$Instituto de F\'{\i}sica, Universidade de S\~{a}o Paulo -- CP 66318, 05314-970 S\~{a}o Paulo, SP, Brazil}}

\vspace{6ex}

{\large \bf Abstract \\}

\vspace{2ex}

\parbox{120mm}
{We show that the infinitesimal generator of the symmetric simple exclusion process, recast as a quantum spin-$\frac{1}{2}$ ferromagnetic Heisenberg model, can be solved by elementary techniques on the complete, complete bipartite, and related multipartite graphs. Some of the resulting infinitesimal generators are formally identical to homogeneous as well as mixed higher spins models. The degeneracies of the eigenspectra are described in detail, and the Clebsch-Gordan machinery needed to deal with arbitrary spin-$s$ representations of the SU($2$) is briefly developed. We mention in passing how our results fit within the related questions of a ferromagnetic ordering of energy levels and a conjecture according to which the spectral gaps of the random walk and the interchange process on finite simple graphs must be equal.

\vspace{2ex}

{\noindent}{\bf Keywords}: \mbox{Simple exclusion process} $\cdot$ \mbox{Heisenberg model} $\cdot$ \mbox{complete graph} $\cdot$ \mbox{Curie-Weiss model} $\cdot$ \mbox{SU(2) algebra} $\cdot$ \mbox{Clebsch-Gordan series}

\vspace{2ex}

{\noindent}{\bf PACS 2010}: 02.50.Ga $\cdot$ 03.65.Fd $\cdot$ 64.60.De

\vspace{2ex}

{\noindent}{\bf Journal ref.}: \href{http://dx.doi.org/10.1088/1751-8113/46/29/295001}{{\it J. Phys. A: Math. Theor.\/} {\bf 46} (2013) 295001 (13pp)}}

\end{center}

\end{titlepage}

\pagestyle{plain}


\section{\label{intro}Introduction}

Exclusion processes, together with the contact process and the Glauber-Ising model, are one of the most fundamental models in the field of nonequilibrium interacting particle systems \cite{liggett}. In physics, exclusion processes are the simplest models that provide nontrivial results on a number of basic issues, such as the relaxation dynamics of an interacting gas towards the thermodynamic equilibrium or the dynamics of shock waves in discrete models for inviscid fluids \cite{spohn}. In the one-dimensional linear chain, simple exclusion processes, either symmetric or asymmetric, under periodic or more general open boundary conditions, have been analyzed and their relationship with other models of interest spanned a wealth of mathematical physics during the last two decades \cite{dhar,varadhan,gwa,dehp,adhr,evans,ptcp19}.

The investigation of exclusion processes on general graphs, however, has received comparatively less attention in the physics literature, despite the fact that in the closely related subject of theoretical magnetism the analysis of models on general graphs has a venerable tradition \cite{temperley,martin,fywu}. Applications of exclusion processes on graphs can be found, e.\,g., in some multilane traffic models and biologically inspired models for intracellular transport and organization \cite{derrida,basu,neri,traffic,ming,ezaki}. In the mathematical literature, otherwise, the study of random walks and exclusion processes on graphs is a hot topic connected with deep results in probability, group theory, harmonic analysis, and  combinatorics \cite{woess,diaconis,saloff}. Unfortunatelly, this literature is difficult to interpret, with possibly useful results hidden behind ramparts of advanced prerequisites, hardcore formalism, and subtle rationale.

In this article we show that the infinitesimal generator of the symmetric simple exclusion process (SSEP) on the complete, complete bipartite, and closely related graphs can be solved by elementary techniques that belong in the toolbox of every trained physicist. We believe that the explicit calculations presented here simplify the understanding of the models and also open some interesting perspectives.

The article is organized as follows. In section~\ref{process} we briefly review the quantum spin formulation for interacting particle systems and display the infinitesimal generator of the SSEP on a graph. In section~\ref{complete}, the SSEP on the complete graph is diagonalized and we characterize its eigenspectrum and the degeneracies of the eigenvalues. The related questions of a ferromagnetic ordering of energy levels and a conjecture on the spectral gaps of the random walk on finite simple graphs are mentioned in this section and then briefly mentioned again later for the other cases treated in the article. The relationship between the spectral gap of the process with its relaxation time is also mentioned. In sections~\ref{bipartite} and \ref{derived} the analyses of section~\ref{complete} are repeated for the SSEP on complete bipartite and complete multipartite graphs. The degeneracies of the eigenspectrum of the SSEP on complete multipartite graphs depend on the outer multiplicities that appear in the Clebsch-Gordan series for arbitrary spin-$s$ representations of the SU($2$), that are derived in the appendix. Section~\ref{derived} also contains some comments on the SSEP on concatenated bipartite graphs, that gives rise to infinitesimal generators formally identical with mixed spins chains. Finally, in section~\ref{summary} we summarize our results and indicate directions for further developments.


\section{\label{process}The SSEP on a graph}

Let $G = (V,E)$ be a finite simple (without loops) undirected connected graph of order $N$ with vertex set $V = \{1, \ldots, N\}$ and edge set $E \subseteq V \times V$. To each $i \in V$ we attach a random variable $\sigma_i$ taking values in $\{-1, +1\}$. If $\sigma_i = -1$ we say that vertex $i$ is empty and if $\sigma_i = +1$ we say that vertex $i$ is occupied by a particle. The state of the system is specified by the configuration $\sigma = (\sigma_1, \ldots, \sigma_N)$ in $\Omega = \{-1,+1\}^{V}$. The SSEP($G$) is the continuous-time Markov jump process that describes the transitions of a set of $n$ itinerant particles, $1 \leq n \leq N$, between the connected vertices of $G$. In the SSEP($G$), each particle chooses, sequentially and at exponentially distributed times, one of its adjacent vertices to jump to provided the target vertex is empty, otherwise the jump attempt fails and the process continues. Clearly, when $n=1$ we have the simple random walk on $G$. When $n \geq 2$, exclusion between particles comes into play and the process becomes more interesting.

We introduce vector spaces in the description of the SSEP($G$) by turning $\Omega$~into $(\mathbb{C}^{2})^{\otimes V}$, $\sigma$ into $|{\sigma}\rangle = |{\sigma_1}\rangle \otimes \cdots \otimes |{\sigma_N}\rangle$, and setting $|{0}\rangle = {0 \choose 1}$ and $|{1}\rangle = {1 \choose 0}$ to identify respectively an empty and an occupied vertex. A little reflection shows that within this vector space scenario the infinitesimal generator of the time evolution of the SSEP($G$) can be written as
\begin{equation}
\label{hpij}
\mathcal{H} = \sum_{i \sim j}\left( 1-\mathcal{P}_{ij} \right),
\end{equation}
where $i \sim j$ stands for pairs of connected vertices of $G$ and $\mathcal{P}_{ij}$ is the operator that transposes the states of vertices $i$ and $j$,
\begin{equation}
\label{pij}
\mathcal{P}_{ij} |{\cdots, \sigma_i, \cdots, \sigma_j, \cdots}\rangle = 
|{\cdots, \sigma_j, \cdots, \sigma_i, \cdots}\rangle.
\end{equation}
Detailed derivations of the evolution operator of the SSEP for the linear chain appear in \cite{adhr,evans,ptcp19}. The derivation for arbitarry graphs follows along the same lines as that for the linear chain, since only the two-body operator $1-\mathcal{P}_{ij}$ really needs to be considered.

As is well known, $\mathcal{P}_{ij}$ can be written in terms of Pauli spin matrices as
\begin{equation}
\label{psisj}
\mathcal{P}_{ij} = \frac{1}{2} (1+\vec{\sigma}_{i} \cdot \vec{\sigma}_{j}) = 
\frac{1}{2} (1 +\sigma^{x}_{i} \sigma^{x}_{j} + \sigma^{y}_{i} \sigma^{y}_{j} + \sigma^{z}_{i} \sigma^{z}_{j}).
\end{equation}
Inserting this $\mathcal{P}_{ij}$ in (\ref{hpij}) gives
\begin{equation}
\label{xxx}
\mathcal{H} = \frac{1}{2} \sum_{i \sim j} (1-\vec{\sigma}_{i} \cdot \vec{\sigma}_{j}).
\end{equation}
We see that $\mathcal{H}$ is, to within a diagonal term, exactly the Hamiltonian of the isotropic Heisenberg spin-$\frac{1}{2}$ quantum ferromagnet over $G$ \cite{mattis}. The ground states of $\mathcal{H}$ have eigenvalue zero and correspond, under a probabilistic normalization, to the stationary states of the process.

Operator (\ref{xxx}) is positive semi-definite and the master equation governing the time evolution of the probability density $P(\sigma,t)$ of observing configuration $\sigma$ at instant $t$ reads $\partial_t P(\sigma,t) = -\mathcal{H}P(\sigma,t)$. One is usually interested in the spectral gap of $\mathcal{H}$, which is the inverse of the leading characteristic time scale of the process related with the time it takes to approach the stationary state. Conservation of particles in the SSEP($G$) implies that $\mathcal{H}$ commutes with the total number of particles operator
\begin{equation}
\label{number}
\mathcal{N} = \frac{1}{2}\sum_{i=1}^{N} (1+\sigma_{i}^{z}) = \frac{N}{2}+\mathcal{S}^{z},
\end{equation}
where $\mathcal{S}^{z}$ is the $z$-axis ``polarization'' operator. It follows that $\mathcal{H}$ is block-diagonal, $\mathcal{H} = \bigoplus_{n}\mathcal{H}_{n}$, with each block $\mathcal{H}_{n}$ acting on its respective invariant subspace $\Omega_{n}$ of dimension $\dim{\Omega_{n}} = {N \choose n} = N!/n!(N-n)!$. The eigenspectrum of $\mathcal{H}$ is also symmetric about $n=N/2$, because it commutes with the ``spin flip'' operator
\begin{equation}
\label{flip}
\mathcal{U} = \prod_{i=1}^{N}\sigma_{i}^{x},
\end{equation}
that transforms particles into holes and vice-versa, $\mathcal{U} |{\sigma_1, \cdots, \sigma_N}\rangle = |{-\sigma_1, \cdots, -\sigma_N}\rangle$, taking a state with $n$ particles into a state with $N-n$ particles. The eigenspectra in the sectors of $n$ and $N-n$ particles are thus identical.

In what follows we investigate operator (\ref{xxx}) on a couple of different graphs and show that some of them can be analyzed by elementary SU($2$) techniques.


\section{\label{complete}The SSEP on the complete graph}

\begin{figure}[t]
\centering
\begin{tabular}{c@{\hspace{4em}}c}
\includegraphics[viewport=124 277 473 561,scale=0.30,angle=-90,clip]{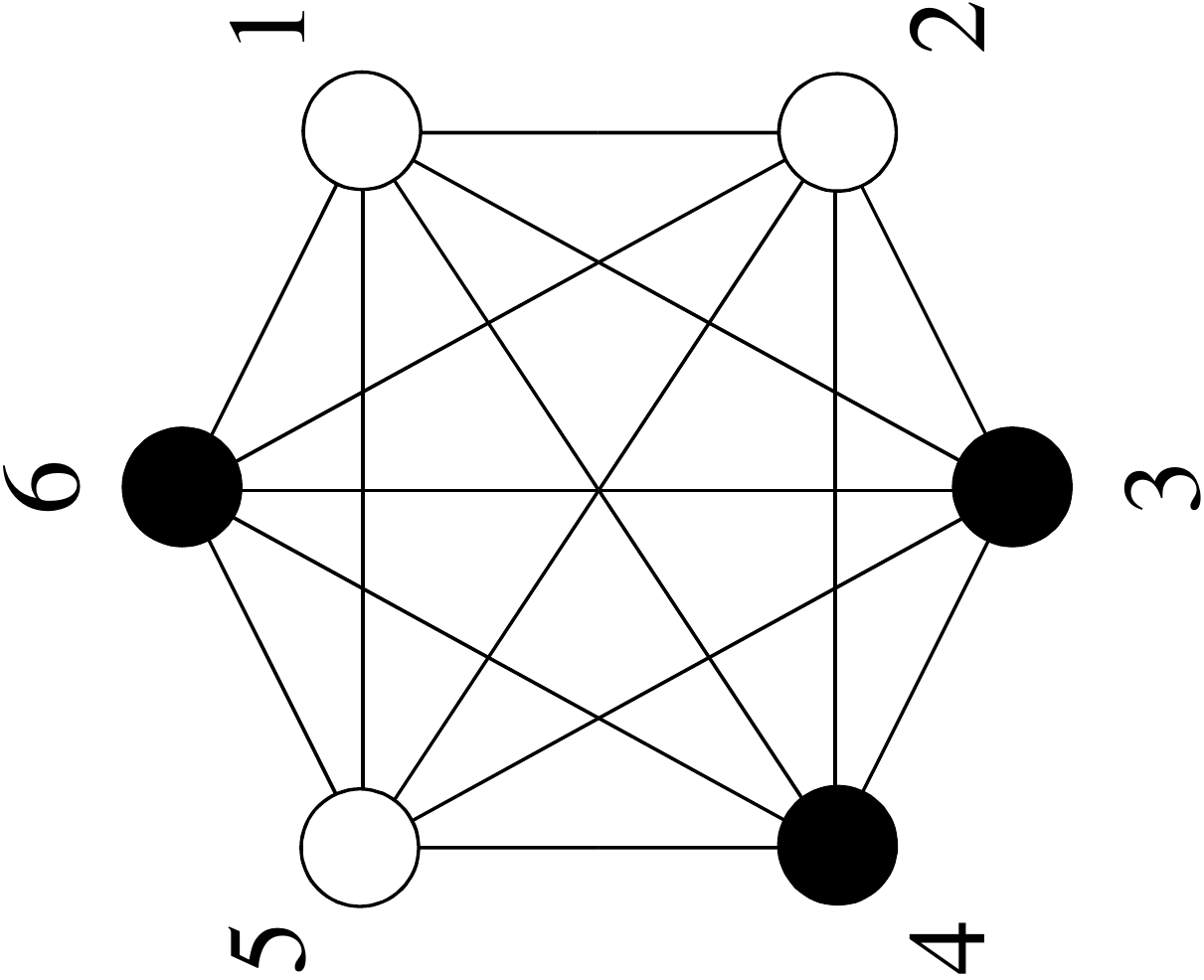} &
\includegraphics[viewport= 35 294 543 548,scale=0.30,angle=-90,clip]{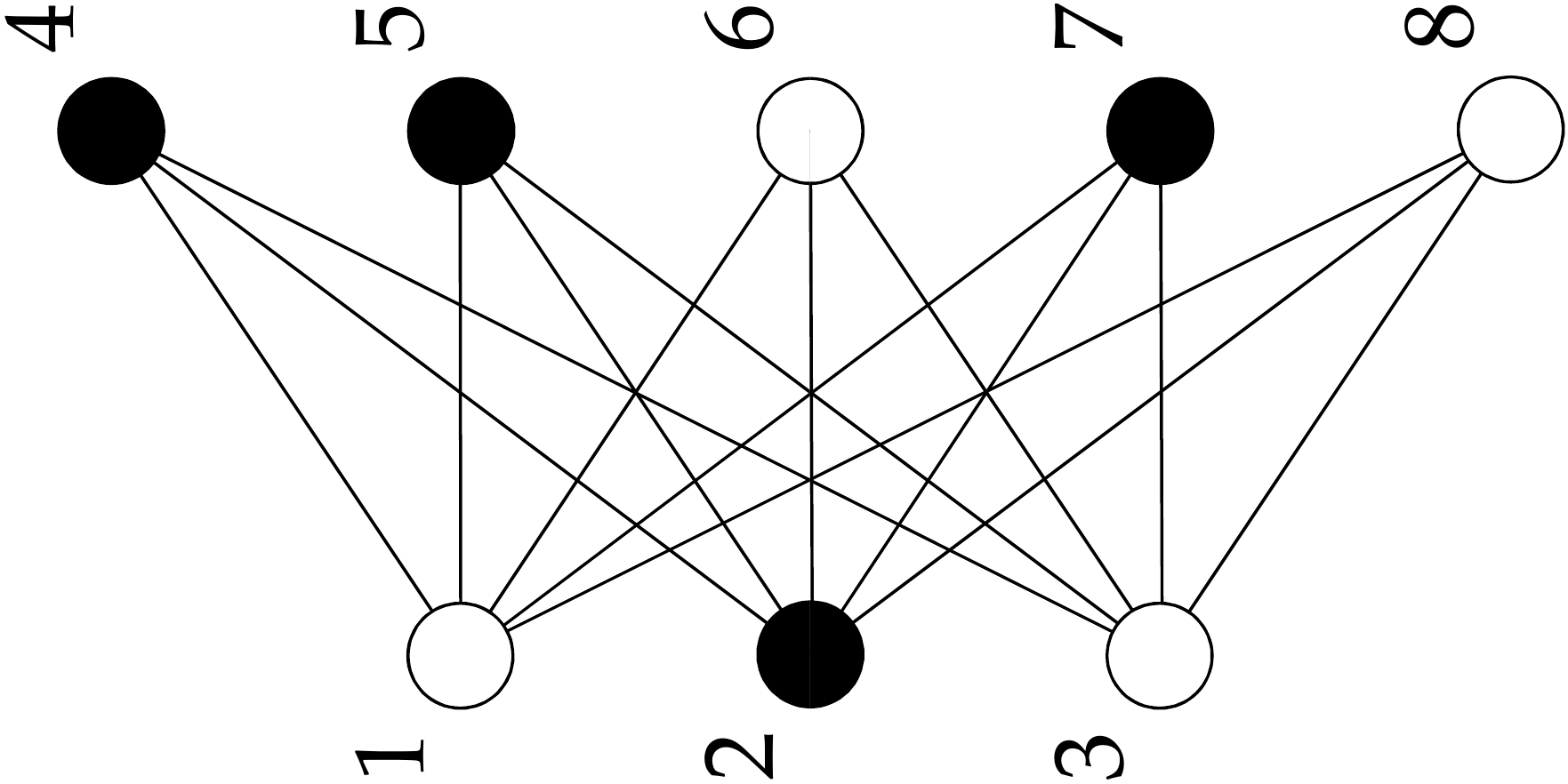}
\end{tabular}
\caption{\label{k6k35}Left: Complete graph $K_{6}$ with $n=3$ particles (black circles) occupying some of its vertices. Right: Complete bipartite graph $K_{3,5}$ occupied by $n=4$ particles.}
\end{figure}

In the complete graph $K_{N}$, every pair of distinct vertices is connected by a unique edge; see figure~\ref{k6k35}. For this graph, the infinitesimal generator (\ref{xxx}) reads
\begin{equation}
\label{hkn}
\mathcal{H} = \frac{1}{2} \sum_{1 \leq i < j \leq N} (1-\vec{\sigma}_{i} \cdot \vec{\sigma}_{j}).
\end{equation}
We can rearrange the nondiagonal part of the summation in (\ref{hkn}) as
\begin{equation}
\sum_{i < j} \vec{\sigma}_{i} \cdot \vec {\sigma}_{j} = 
\frac{1}{2}\sum_{i < j} \vec{\sigma}_{i} \cdot \vec {\sigma}_{j} + \frac{1}{2}\sum_{i > j} \vec{\sigma}_{i} \cdot \vec {\sigma}_{j} =\frac{1}{2}\Big(\sum_{i} \vec{\sigma}_{i}\Big)\Big(\sum_{j} \vec{\sigma}_{j}\Big)-\frac{1}{2}\sum_{i} \vec{\sigma}_{i}^{2},
\end{equation}
where in the last passage we added and subtracted the diagonal term $\frac{1}{2}\sum_{i = j} \vec{\sigma}_{i} \cdot \vec {\sigma}_{j}$ and factored the resulting unrestricted double sum. Since $\sum_{1 \leq i < j \leq N} 1 = N(N-1)/2$---this term is just the total number of edges of the graph, $\sum_{i \sim j} 1 = | E |$---and $\vec{\sigma}_{i}^{2} = 3_{i}$, we eventually arrive at
\begin{equation}
\label{weiss}
\mathcal{H} = \frac{N}{2}\Big(\frac{N}{2}+1\Big)
 -\Big(\frac{1}{2}\sum_{i=1}^{N}\vec{\sigma}_{i}\Big)^2.
\end{equation}
This $\mathcal{H}$ is but the Curie-Weiss version of the spin-$\frac{1}{2}$ ferromagnetic Heisenberg model without the overall multiplicative $1/N$ term usually included to keep the energy per spin an intensive quantity, since we are not doing any thermodynamics here \cite{kastner}. In the basis simultaneously diagonal in the total spin squared operator
\begin{equation}
\label{square}
\vec{\mathcal{S}}^2 = \Big( \frac{1}{2} \sum_{i=1}^{N} \vec{\sigma}_{i} \Big)^2
\end{equation}
with eigenvalues $S(S+1)$, $S = S_{\rm min}, S_{\rm min}+1, \ldots, N/2$, where $S_{\rm min}=0$ or $1/2$ depending whether $N$ is even or odd, and in the total $z$-axis component $\mathcal{S}^{z}$ defined in (\ref{number}) with eigenvalues $M = -S, -S+1, \ldots, +S$, the eigenvalues of $\mathcal{H}$ read
\begin{equation}
\label{esm}
E_{N}(S,M) = \frac{N}{2}\Big(\frac{N}{2}+1\Big) - S(S+1),
\end{equation}
i.e., $E_{N}(S,M) = E_{N}(S)$, independent on $M$. The degeneracy of $E_{N}(S)$ is given by $g_{N}(S)=$ $(2S+1) \times d_{1/2}(N,S)$, where the factor $2S+1$ comes from the degeneracy in the $\mathcal{S}^{z}$ values of rotationally invariant operators like $\mathcal{H}$, and the $d_{1/2}(N,S)$ comes from the fact that there exists many possible combinations of the $N$ elementary spins summing up to a definite value of $S$. This last factor is given by the outer multiplicity of the irreducible representation $\mathcal{D}^{(S)}$ appearing in the Clebsch-Gordan series
\begin{equation}
\label{clebsch}
[\mathcal{D}^{(1/2)}]^{\otimes N} = 
\bigoplus_{S=S_{\rm min}}^{N/2} d_{1/2}(N,S) \mathcal{D}^{(S)},
\end{equation}
and can be shown to be given by (cf. appendix)
\begin{equation}
\label{deg}
d_{1/2}(N,S) = {N \choose \frac{1}{2}N+S} - {N \choose \frac{1}{2}N+S+1}.
\end{equation}

We see from Eqs. (\ref{esm}) and (\ref{deg}) that $E_{N}(S=N/2) = 0$ with a $(N+1)$-fold degeneracy. These values have a simple interpretation: the SSEP($G$) has a zero eigenvalue on each of its $N+1$ sectors of total particle number $n = N/2+M = 0, 1, \ldots, N$. That the stationary states of $\mathcal{H}$ occur in the sectors of $S=N/2$ is just another statement of the well known fact that the ground states of ferromagnetic Heisenberg models have maximum possible total $S$. The right eigenvectors corresponding to the zero eigenvalues are the stationary states of the process, explicitly given by
\begin{equation}
\label{states}
|{\Phi_{0}^{N}(n)}\rangle = {N \choose n}^{\!-1} \! \sum_{1 \leq i_{1} < i_{2} < \cdots < i_{n} \leq N} |{1_{i_{1}}, 1_{i_{2}}, \cdots, 1_{i_{n}}}\rangle;
\end{equation}
notice the probabilistic normalization of $|{\Phi_{0}^{N}(n)}\rangle$, not the quantum-mechanical one. The summation in (\ref{states}) runs over all combinations of the $n$ particle positions $i_{1}$, $i_{2}$, \ldots, $i_{n}$ among the $N$ available vertices of the graph.

For processes that conserve the total number of particles like the SSEP($G$), a basis diagonal in $n$ is more useful. In the $|{S,M}\rangle$ basis, each invariant subspace $\Omega_{n}$ of fixed $n = 0, 1, \ldots, N$ is spanned by the states with $M=-N/2+n$ fixed and $| -N/2+n | \leq S \leq N/2$, with the given $|{S,M=-N/2+n}\rangle$ states within $\Omega_{n}$ bearing their original multiplicity $d_{1/2}(N,S)$. This completely characterizes the eigenspectrum of $\mathcal{H}$ in each of its invariant subspaces.

Tables \ref{tab:ek8s} and \ref{tab:ek8n} illustrate the SSEP($K_{N}$) in the concrete case of $N=8$. The eigenvalues of SSEP($K_{8}$) and their degeneracies appear in table~\ref{tab:ek8s}. The eigenspectrum in terms of the total number of particles appears in table~\ref{tab:ek8n} and is clearly symmetric about $n=N/2$, as we anticipated in section~\ref{process}.

The spectral gap $\Delta_{N}$ of $\mathcal{H}$ is given by the smallest nonzero eigenvalue of $\mathcal{H}$, and is related with the characteristic time $\tau_{N}$ it takes for the process to exponentially decay to its stationary state by $\tau_{N}^{-1} = \Delta_{N}$. For the SSEP($K_{N}$), $\Delta(K_{N}) = E_{N}(N/2-1) = N$ is the same in every invariant sector of constant particle number of the process, except in the one-dimensional sectors of $n=0$ and $n=N$, for which there is no gap at all. That $\Delta(K_{N}) = N$ hints at the fact that the characteristic time $\tau_{N}$ scales with the system size as $N\tau_{N} = N\Delta(K_{N})^{-1} = 1$, i.e., the interacting particle system relaxes to its stationary state after just one step, irrespective of $N$. This has to do with the fact that on $K_{N}$ any vertex can be reached from any other one through a single jump.

It is well known that antiferromagnetic models over bipartite lattices have the ground state in the subspace of least possible total spin $S$, with the lowest-lying eigenvalues in the subspaces of $S$ obeying an antiferromagnetic ordering of energy levels, $E_{0}(S') > E_{0}(S)$ for $S' > S$. This is the contents of the Lieb-Mattis theorem \cite{lieb-mattis}. For ferromagnetic models, otherwise, the state of minimum energy occurs in the subspace of maximum total $S$, and there is no {\it a priori\/} rigorously established ordering for the eigenvalues with $S$. Recently, however, a ferromagnetic analog of the Lieb-Mattis theorem was developed for some ferromagnetic SU($2$)-invariant quantum spin models \cite{nachter}. As it can be seen from (\ref{esm}), the energy levels of the SSEP($K_{N}$) are monotone decreasing in $S$, $E_{N}(S') < E_{N}(S)$ whenever $S' > S$, thus observing a ``ferromagnetic ordering of energy levels.''  All models analyzed in this article observe this type of ordering of eigenvalues. Despite the ubiquity of this type of ordering among ferromagnetic models, counterexamples to this ordering property were found recently for graph topologies as simple as the cycle graph $C_{N}$ (the one-dimensional periodic lattice) with an even number of vertices \cite{starr}.

The ferromagnetic ordering of energy levels in the SSEP($K_{N}$) is also akin to the so-called Aldous' spectral gap conjecture, according to which the gap of the single particle random walk ($n=1$) should be equal to the gap of the interchange process ($n=N$) on any finite simple graph \cite{aldous}. This conjecture spawned some original results in probability and mathematical physics, mostly over the last decade, and was proved in general only recently through a {\it m\'{e}lange\/} of group-theoretical, probabilistic, and combinatorial arguments \cite{caputo}.

\begin{table}[t]
\centering
\caption{\label{tab:ek8s}Eigenspectrum of the SSEP($G$) on the complete graph $K_{8}$. The degeneracies $g_{N}(S)$ are given as $(2S+1) \times d_{1/2}(N,S)$.  Notice that $\sum_{S}g_{N}(S) = 2^{8}$, as it should be.}
\medskip
\begin{tabular}{cccccc}
\hline\hline
   $S$     &   0  &   1  &   2  &  3  &  4  \\ 
\hline
$E_{N}(S)$ &  20  &  18  &  14  &  8  &  0  \\ 
$g_{N}(S)$ & $1 \times 14$ & $3 \times 28$ & $5\times 20$ & $7\times 7$ & $9\times 1$ \\ 
\hline\hline
\end{tabular}
\end{table}

\begin{table}[t]
\centering
\caption{\label{tab:ek8n}Characterization of the invariant subspaces $\Omega_{n}$ of the SSEP($K_{8}$). The multiplicities of the $|{S,M=-N/2+n}\rangle$ states within each $\Omega_{n}$ are given in the last column as $(S^{d_{1/2}(N,S)})$.}
\medskip
\begin{tabular}{ccc}
\hline\hline
$n=\frac{1}{2}N+M$ & $\dim{\Omega_{n}}$ & ($S^{d_{1/2}(N,S)}$) \\
\hline
 0 &  1 & $(4^{1})$ \\ 
 1 &  8 & $(3^{7})(4^{1})$ \\ 
 2 & 28 & $(2^{20})(3^{7})(4^{1})$ \\ 
 3 & 56 & $(1^{28})(2^{20})(3^{7})(4^{1})$ \\ 
 4 & 70 & $(0^{14})(1^{28})(2^{20})(3^{7})(4^{1})$ \\ 
 5 & 56 & $(1^{28})(2^{20})(3^{7})(4^{1})$ \\ 
 6 & 28 & $(2^{20})(3^{7})(4^{1})$ \\
 7 &  8 & $(3^{7})(4^{1})$ \\ 
 8 &  1 & $(4^{1})$ \\
\hline\hline
\end{tabular}
\end{table}


\section{\label{bipartite}The SSEP on the complete bipartite graph}

The complete bipartite graph $K_{N_{1},N_{2}}$ is the simple undirected graph with partitioned vertex set $V = V_{1} \cup V_{2}$ with $|{V_{i}}| = N_{i}$, $i=1,2$, and $V_{1} \cap V_{2} = \varnothing$ such that every vertex in $V_{1}$ is connected to every vertex in $V_{2}$ by a unique edge; see figure~\ref{k6k35}. For this graph,
\begin{equation}
\label{s1s2}
\mathcal{H} = \frac{1}{2} \sum_{i_{1} \in V_{1}} \sum_{i_{2} \in V_{2}}
(1 - \vec{\sigma}_{i_{1}} \cdot \vec{\sigma}_{i_{2}}) = 
\frac{1}{2}N_{1}N_{2} - 2\vec{\mathcal{S}}_{1} \cdot \vec{\mathcal{S}}_{2},
\end{equation}
where the operators $\vec{\mathcal{S}}_{1}$ and $\vec{\mathcal{S}}_{2}$ are given by
\begin{equation}
\label{subtotal}
\vec{\mathcal{S}}_{1} = \frac{1}{2}\sum_{i_{1} \in V_{1}}\vec{\sigma}_{i_{1}}, \quad
\vec{\mathcal{S}}_{2} = \frac{1}{2}\sum_{i_{2} \in V_{2}}\vec{\sigma}_{i_{2}}.
\end{equation}
It is a matter of simple algebra to demonstrate that the $\vec{\mathcal{S}}_{i}$ obey $\vec{\mathcal{S}}_{i} \times \vec{\mathcal{S}}_{i} = \I \vec{\mathcal{S}}_{i}$, $i=1,2$, being thus legitimate spin operators. The magnitude of the spin $\vec{\mathcal{S}}_{i}$ is $S_{i} = N_{i}/2$.

Notice that in representing the occupation state of $V_{i}$ by a state of $\vec{\mathcal{S}}_{i}$ indexed by the value of its $\mathcal{S}_{i}^{z}$ component through the relation $n_{i}=N_{i}/2+m_{i}$, $m_{i}= -N_{i}/2$, $-N_{i}/2+1$, \ldots, $+N_{i}/2$, we have promoted a reduction of the dimension of the configuration space associated with $V_{i}$ from $2^{N_{i}}$ to $N_{i}+1$. This reduction comes from lumping equivalent configurations obtained by permutations of the particles among the vertices of $V_{i}$ into a single representative state. The result is that the $2^{N}$-dimensional original problem can be treated as a $(N_{1}+1)(N_{2}+1)$-dimensional problem as far as the determination of the eigenspectrum is concerned. If the eigenstates of (\ref{s1s2}) become needed, e.\,g., to calculate correlation functions or block entropies, one must reconstruct them from the original $2^{N}$ states by appropriate combinations of permutations.

In terms of the total spin $\vec{\mathcal{S}} = \vec{\mathcal{S}}_{1} + \vec{\mathcal{S}}_{2}$, we have $2\vec{\mathcal{S}}_{1} \cdot \vec{\mathcal{S}}_{2} = \vec{\mathcal{S}}^{2} - \vec{\mathcal{S}}_{1}^{2} - \vec{\mathcal{S}}_{2}^{2}$. In the basis diagonal in the complete set of commuting operators $\vec{\mathcal{S}}_{1}^{2}$, $\vec{\mathcal{S}}_{2}^{2}$, $\vec{\mathcal{S}}^{2}$, and $\mathcal{S}^{z} = \mathcal{S}_{1}^{z} + \mathcal{S}_{2}^{z}$, the eigenvalues of (\ref{s1s2}) are given by
\begin{equation}
\label{es1s2}
E_{N_{1},N_{2}}(S_{1},S_{2},S,M) =
\frac{1}{2}N_{1}N_{2}+S_{1}(S_{1}+1)+S_{2}(S_{2}+1)-S(S+1),
\end{equation}
or, in more compact form, by
\begin{equation}
\label{e-compact}
E_{N}(S) = (N/2-S)(N/2+S+1),
\end{equation}
with $|{S_{1}-S_{2}}| \leq S \leq S_{1}+S_{2}$ and $|{M}| \leq S$. The number of particles in the system is given by $n=N/2+M$, as before.

For each of the $\min \{ 2S_{1}+1, 2S_{2}+1 \}$ values of $S$, $E_{N}(S)$ is $2S+1$ degenerate due to its independence on $M$. Overall, $M$ is in the range $-|{S_{1}-S_{2}}| \leq M \leq S_{1}+S_{2}$, and for any given $M$ we have $\max \{|{M}|,|{S_{1}-S_{2}}|\} \leq S \leq S_{1}+S_{2}$. The lowest eigenvalue of (\ref{s1s2}) lies in the sector of maximum $S = S_{1}+S_{2} = N_{1}/2+N_{2}/2 = N/2$---as expected for a ``ferromagnetic'' model---, with a $N+1$ degeneracy ($M = -N/2, -N/2+1, \ldots, +N/2$) associated with the $N+1$ stationary states of the process, one within each invariant sector of constant number of particles ($n = N/2+M = 0, 1, \ldots, N$). The steady states are given by the same $|{\Phi_{0}^{N}(n)}\rangle$ as in (\ref{states}).

The spectral gap of the process is given by $\Delta(K_{N_{1},N_{2}}) = E_{N}(N/2-1) = N+2$, and like the gap of the SSEP($K_{N}$) is the same in every invariant sector of constant particle number. It is also clear from (\ref{es1s2}) or (\ref{e-compact}) that the energy levels observe the ferromagnetic ordering property mentioned in section~\ref{complete}, namely, $E_{N}(S') < E_{N}(S)$ whenever $S' > S$, providing yet another example of such systems \cite{nachter,starr}.


\section{\label{derived}The SSEP on multipartite graphs}

The cases analyzed so far lead naturally to the SSEP on generalized multipartite graphs. In particular, two types of multipartite graphs are of interest: complete mutipartite graphs and concatenated (chained) bipartite graphs. Although the SSEP on this second type of graphs cannot be solved by elementary techniques in general---actually, most of them cannot be exactly solved at all---, they give rise to infinitesimal generators that may appeal in other modeling circumstances.


\subsection{The complete multipartite graph}

\begin{figure}[t]
\centering
\includegraphics[viewport=147 253 455 588,scale=0.40,angle=-90,clip]{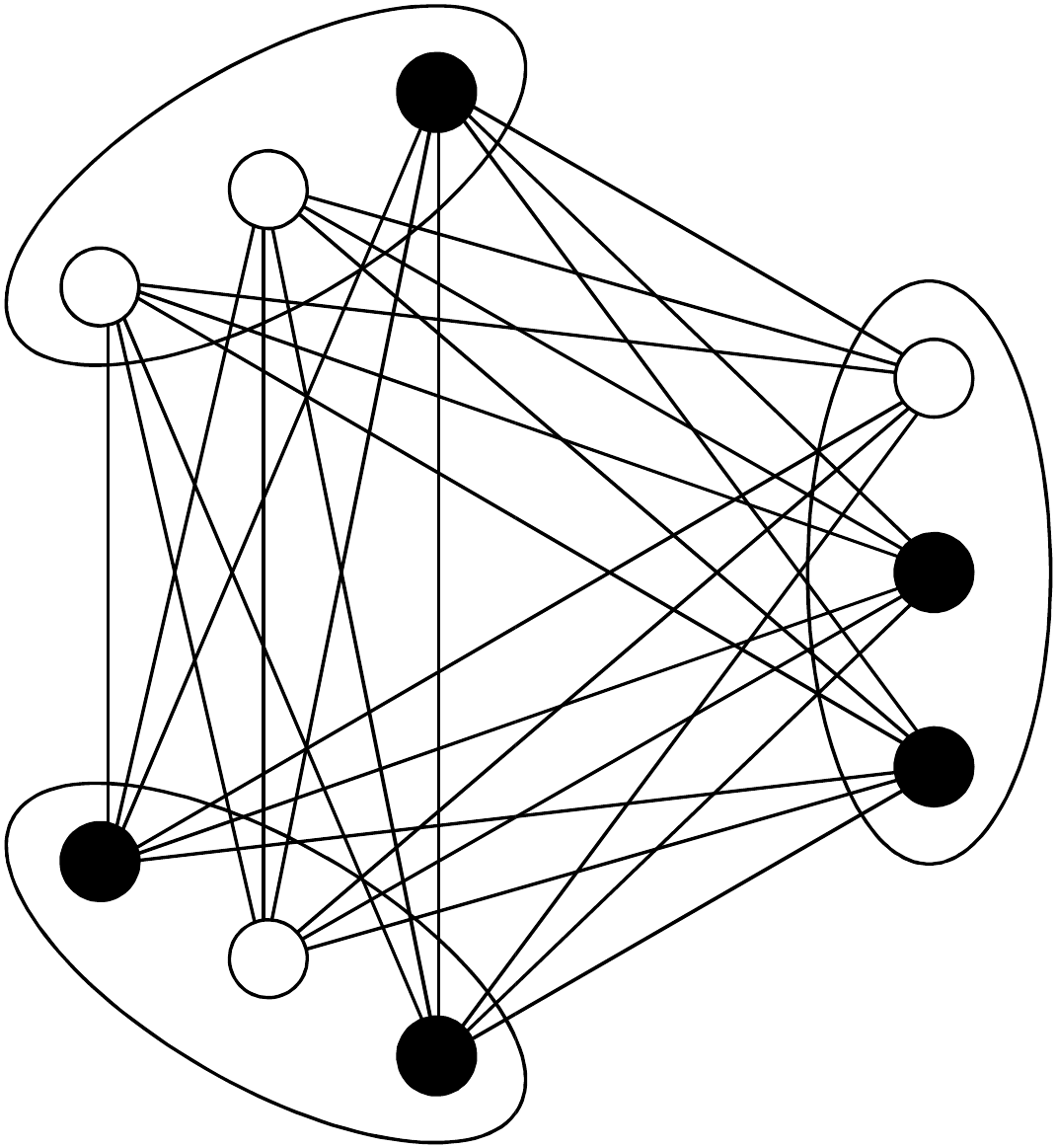}
\caption{\label{kkk}The $3$-regular complete multipartite graph $K_{3,3,3} = K_{3}^{3}$ occupied by $n=5$ particles (black circles). Clearly, $K_{1,\ldots,1} = K_{1}^{N} = K_{N}$, the complete graph considered in section~\ref{complete}. If we interpret the sets $V_{i}$ as ``urns'' holding up to $Q_{i}$ particles each, the SSEP($K_{Q_{1},\cdots,Q_{N}}$) becomes the symmetric partial exclusion process on the complete graph.}
\end{figure}

The complete multipartite graph $K_{Q_{1},\cdots,Q_{N}}$ is the simple undirected graph with partitioned vertex set $V=V_{1} \cup \cdots \cup V_{N}$ with $|{V_{i}}| = Q_{i}$, $i = 1, \ldots, N$, and $V_{i} \cap V_{j} = \varnothing$ for $i \neq j$ such that every two vertices from different sets $V_{i}$ and $V_{j}$ are adjacent. When $Q_{1} = \cdots = Q_{N} = Q$, we have the $Q$-regular complete multipartite graph $K_{Q}^{N}$; see figure~\ref{kkk}.

Following our previous approach, we associate to each disjoint subset $V_{k}$ a spin-$Q_{k}/2$ operator $\vec{\mathcal{S}}_{k}$ acting on its own subspace of dimension $Q_{k}+1$ given by
\begin{equation}
\label{skvk}
\vec{\mathcal{S}}_{k} = \frac{1}{2} \sum_{i_{k} \in V_{k}} \vec{\sigma}_{i_{k}}.
\end{equation}
In terms of these spin operators, the infinitesimal generator of the SSEP on the complete multipartite graph $K_{Q_{1},\cdots,Q_{N}}$ reads
\begin{equation}
\label{hknq}
\mathcal{H} = \frac{1}{4}\Big(\sum_{i=1}^{N}Q_{i}\Big)^{2}
-\frac{1}{4}\sum_{i=1}^{N} Q_{i}^{2}
-\Big(\sum_{i=1}^{N}\vec{\mathcal{S}}_{i}\Big)^{2}
+\sum_{i=1}^{N}\vec{\mathcal{S}}_{i}^{2}.
\end{equation}

On the $Q$-regular complete multipartite graph $K_{Q}^{N}$, all $\vec{\mathcal{S}}_{i}$ are equivalent spin-$Q/2$ operators. In this case, taking into account that $\vec{\mathcal{S}}_{i}^{2} = \frac{1}{2}Q(\frac{1}{2}Q+1)$, the infinitesimal generator (\ref{hknq}) of the SSEP($K_{Q}^{N}$) becomes
\begin{equation}
\label{nqweiss}
\mathcal{H} = \frac{1}{2}NQ\Big(\frac{1}{2}NQ+1\Big)
-\Big(\sum_{i=1}^{N}\vec{\mathcal{S}}_{i}\Big)^{2}.
\end{equation}
This operator is formally identical with the Hamiltonian of a quantum spin-$Q/2$ Curie-Weiss model and can be analyzed along the same lines as the spin-$\frac{1}{2}$ operator (\ref{deg}) in section~\ref{complete}. In terms of the total spin operator $\vec{\mathcal{S}} = \sum_{i} \vec{\mathcal{S}}_{i}$, the eigenvalues of (\ref{nqweiss}) can be read off immediately as
\begin{equation}
\label{enq}
E_{Q}^{N}(S,M) = \frac{1}{2}NQ\Big(\frac{1}{2}NQ+1\Big)-S(S+1),
\end{equation}
with $S = S_{\rm min}, S_{\rm min}+1, \ldots, NQ/2$ and $M = -S, -S+1, \ldots, +S$, where $S_{\rm min}=1/2$ if $Q$ and $N$ are both odd and $S_{\rm min}=0$ otherwise. Within each sector of fixed number of particles $n=NQ/2+M$, the values of $S$ range in the interval $|{M}| \leq S \leq NQ/2$. The degeneracies associated with the eigenvalues (\ref{enq}) are given by $g_{N}(S) = (2S+1) \times d_{Q/2}(N,S)$, where now the outer multiplicities $d_{Q/2}(N,S)$ determining the degeneracies in the $S$ values are given by (cf. appendix)
\begin{equation}
\label{dqbb}
d_{Q/2}(N,S) = b_{Q/2}(N,S) - b_{Q/2}(N,S+1),
\end{equation}
where the coefficients $b_{Q/2}(N,M)$ are given by
\begin{equation}
\label{bqn}
b_{Q/2}(N,M) =
\sum_{k \geq 0} (-1)^{k}
{N \choose k} {(\frac{1}{2}Q+1)N + M -(Q+1)k -1 \choose \frac{1}{2}QN + M -(Q+1)k},
\end{equation}
where the summation runs over $k$ as long as the summing terms are non-null. Equations~(\ref{dqbb})--(\ref{bqn}) are analogous to (\ref{deg}) and, indeed, they reduce to it when $Q=1$. Coefficient (\ref{bqn}) also corresponds to the dimension of the invariant subspace $\Omega_{M}$ of fixed $M=|{-NQ/2+n}|$; notice that $\dim{\Omega_{M}} = \dim{\Omega_{-M}}$. As an example, the eigenspectrum of the SSEP($K_{3}^{4}$) and its degeneracies appears in table~\ref{tab:ek34n}.

The same observations made for the eigenspectrum (\ref{esm}) hold here. The zero eigenvalue in \ref{enq}) occurs in the sector of $S=NQ/2$ with a $NQ+1$-fold degeneracy, corresponding to the $NQ+1$ stationary states of the process, one within each subspace of constant particle number. The spectral gap $\Delta(K_{Q}^{N}) = NQ$ of the process is also the same within each invariant sector of constant particle number. Finally, it is clear from (\ref{enq}) that the eigenvalues observe the ferromagnetic ordering $E_{Q}^{N}(S') < E_{Q}^{N}(S)$ if $S' > S$. In fact, the SSEP($K_{Q}^{N}$) and the SSEP($K_{N}$) differ only by the total spin associated with each vertex set, barring the additional degeneracies induced by the permutational equivalence of states within each vertex set that we briefly discussed in section~\ref{bipartite}. Notice that if we interpret the vertices $V_{i}$ of $K_{Q_{1},\cdots,Q_{N}}$ as ``urns'' that can hold up to $Q_{i}$ particles before enforcing exclusion, the SSEP($K_{Q}^{N}$) becomes the symmetric partial exclusion process on the complete graph. In this interpretation the permutation degeneracy mentioned before becomes a nonissue, at the expense of more complicated commutation relations between the operators involved. The partial exclusion process on the linear chain has been analyzed in \cite{schutz,tailleur}, which display a host of techniques and results relevant to our subject.

\begin{table}[t]
\centering
\caption{\label{tab:ek34n}Characterization of the invariant subspaces $\Omega_{n}$ for the SSEP($K_{3}^{4}$). The dimensionality of $\Omega_{n}$ is given by the coefficient $b_{Q/2}(N,M=|{-NQ/2+n}|)$ given in (\ref{bqn}). The multiplicities of the $|{S,M=-NQ/2+n}\rangle$ states within each $\Omega_{n}$ are given in the last column as $(S^{d_{Q/2}(N,S)})$. The additional multiplicities coming from the permutation equivalent states within each set $V_{Q_{i}}$, $i=1, \ldots, N$, are not accounted for in this table.}
\medskip
\begin{tabular}{ccc}
\hline\hline
$n=\frac{1}{2}NQ+M$ & $\dim{\Omega_{n}}$ & ($S^{d_{Q/2}(N,S)}$) \\
\hline
 0  &  1 & $(6^{1})$ \\ 
 1  &  4 & $(5^{3})(6^{1})$ \\ 
 2  & 10 & $(4^{6})(5^{3})(6^{1})$ \\ 
 3  & 20 & $(3^{10})(4^{6})(5^{3})(6^{1})$ \\ 
 4  & 31 & $(2^{11})(3^{10})(4^{6})(5^{3})(6^{1})$ \\ 
 5  & 40 & $(1^{9})(2^{11})(3^{10})(4^{6})(5^{3})(6^{1})$ \\ 
 6  & 44 & $(0^{4})(1^{9})(2^{11})(3^{10})(4^{6})(5^{3})(6^{1})$ \\ 
 7  & 40 & $(1^{9})(2^{11})(3^{10})(4^{6})(5^{3})(6^{1})$ \\ 
 8  & 31 & $(2^{11})(3^{10})(4^{6})(5^{3})(6^{1})$ \\ 
 9  & 20 & $(3^{10})(4^{6})(5^{3})(6^{1})$ \\ 
 10 & 10 & $(4^{6})(5^{3})(6^{1})$ \\ 
 11 &  4 & $(5^{3})(6^{1})$ \\ 
 12 &  1 & $(6^{1})$ \\ 
\hline\hline
\end{tabular}
\end{table}


\subsection{Concatenated bipartite graphs in a chain}

\begin{figure}[t]
\centering
\includegraphics[viewport=182 114 413 727,scale=0.40,angle=-90,clip]{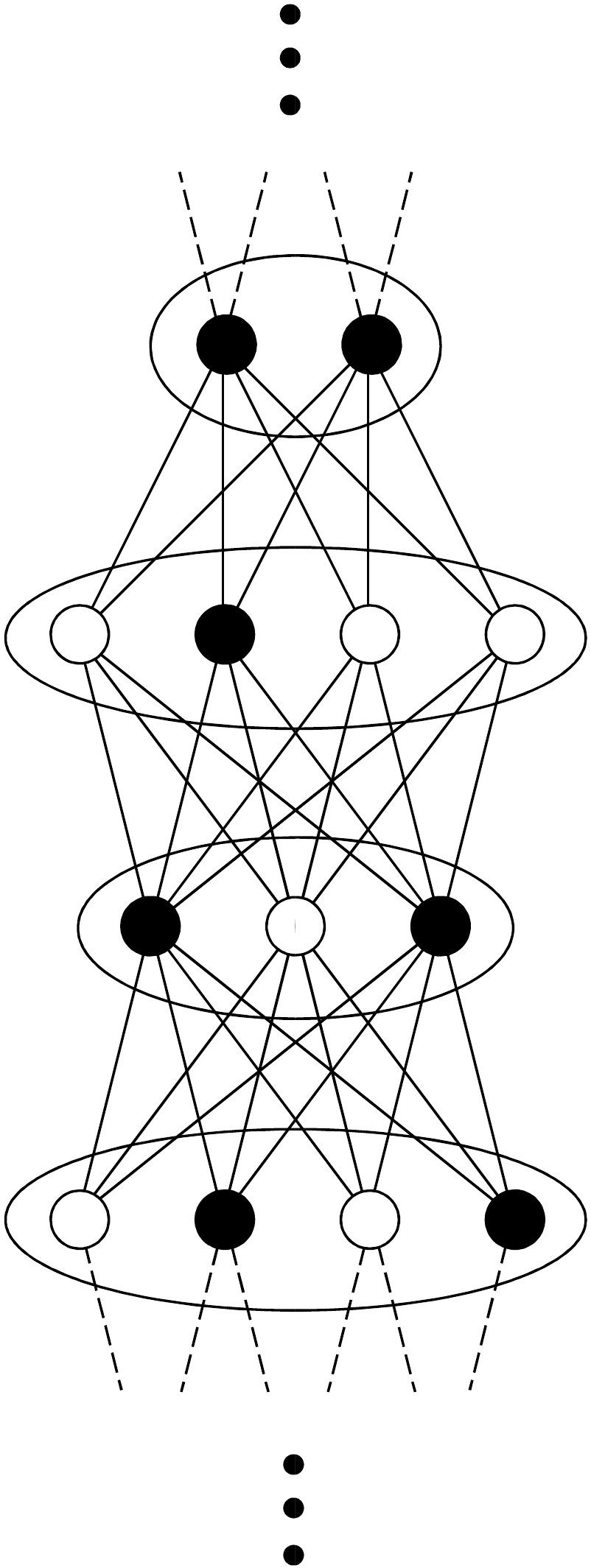}
\caption{\label{k43k34}Concatenated bipartite graphs in a chain. The segment shown depicts graphs $K_{43}$, $K_{34}$, and $K_{42}$ linked together such that every vertex is linked to every other vertex in the ``neighboring'' vertices sets. In general, the edges set of the composite graph under open boundary condition is given by $E_{\rm open} = (V_{Q_{1}} \times V_{Q_{2}}) \cup (V_{Q_{2}} \times V_{Q_{3}}) \cup \cdots \cup (V_{Q_{N}} \times V_{Q_{N+1}})$; under closed boundary condition the edges set becomes $E_{\rm closed} = E_{\rm open} \cup (V_{Q_{N+1}} \times V_{Q_{1}})$.}
\end{figure}

If we concatenate $N$ bipartite graphs, we obtain a graph like the one depicted partly in figure~\ref{k43k34}. We shall denote this graph as $K_{Q_{1},Q_{2}} \times K_{Q_{2},Q_{3}} \times \cdots \times K_{Q_{N},Q_{N+1}}$. Under periodic boundary conditions there are additional edges between $K_{Q_{N},Q_{N+1}}$ and $K_{Q_{1},Q_{2}}$, and in this case we must have $Q_{N+1}=Q_{1}$, otherwise it is an open chain. For this graph, under open boundary conditions $\mathcal{H}$ reads
\begin{equation}
\label{mixed}
\mathcal{H} = \frac{1}{2} \sum_{i=1}^{N}Q_{i}Q_{i+1}-
2\sum_{i=1}^{N}\vec{\mathcal{S}}_{i} \cdot \vec{\mathcal{S}}_{i+1},
\end{equation}
where the spin operators $\vec{\mathcal{S}}_{i}$ are given as in (\ref{skvk}). Operator (\ref{mixed}) is equivalent to the Hamiltonian of a one-dimensional ferromagnetic Heisenberg model of mixed spins $S_{1}=Q_{1}/2$, \ldots, $S_{N+1}=Q_{N+1}/2$, with each bipartite graph $K_{Q_{i},Q_{i+1}}$ of the chain corresponding to a ``unit cell.''

If all $Q_{i}$ are equal, then a simple spin-wave analysis shows that the low-lying eigenspectrum just above the stationary state has the form $E_{N} \propto 2\sin^{2}(\pi/N)$, with an asymptotic behavior $E_{N} \sim 2\pi^{2}N^{-2}$ for $N \gg 1$ \cite{mattis}. In this case, the relation between the relaxation time scale and the spectral gap becomes $\tau \sim N^{2}$, typical of diffusive behavior. It is well known that the SSEP displays this type of dispersion relation, where the dependence on the number of particles and on the spin magnitude affects only prefactors, not the dependence on $N^{2}$ \cite{liggett,spohn}.

If some or all $Q_{i}$ are different, then we have a full-fledged mixed-spins operator. It has been proved that the eigenspectrum of the mixed-spins chain (\ref{mixed}) displays the ferromagnetic ordering of energy levels; actually, this property was first demonstrated for quantum spins chains like (\ref{mixed}) under open boundary conditions \cite{nachter}. While mixed-spins Hamiltonians of interest in the theory of magnetism are usually antiferromagnetic, operator (\ref{mixed}) is always ferromagnetic \cite{note}. We thus expect that a modified spin-wave analysis already succesful in the more complicated cases of antiferromagnetic or competing interactions \cite{mixspin} shall work in the analysis of (\ref{mixed}) as well. This provides an interesting avenue for further investigations.

\section{\label{summary}Summary and outlook}

\begin{figure}[t]
\centering
\begin{tabular}{c@{\hspace{4em}}c}
\includegraphics[viewport=120 246 451 597,scale=0.30,angle=-90,clip]{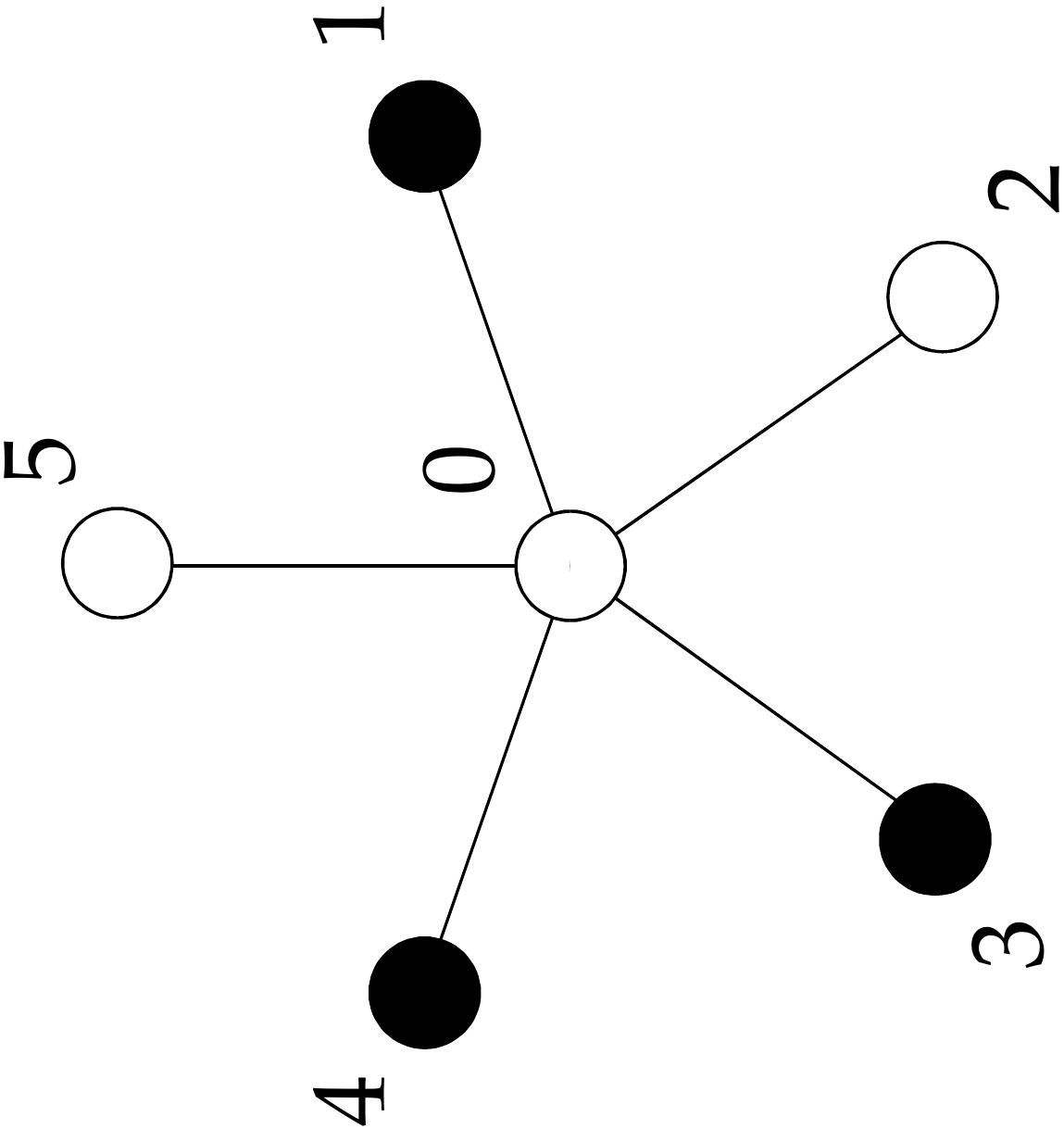} &
\includegraphics[viewport=120 246 451 597,scale=0.30,angle=-90,clip]{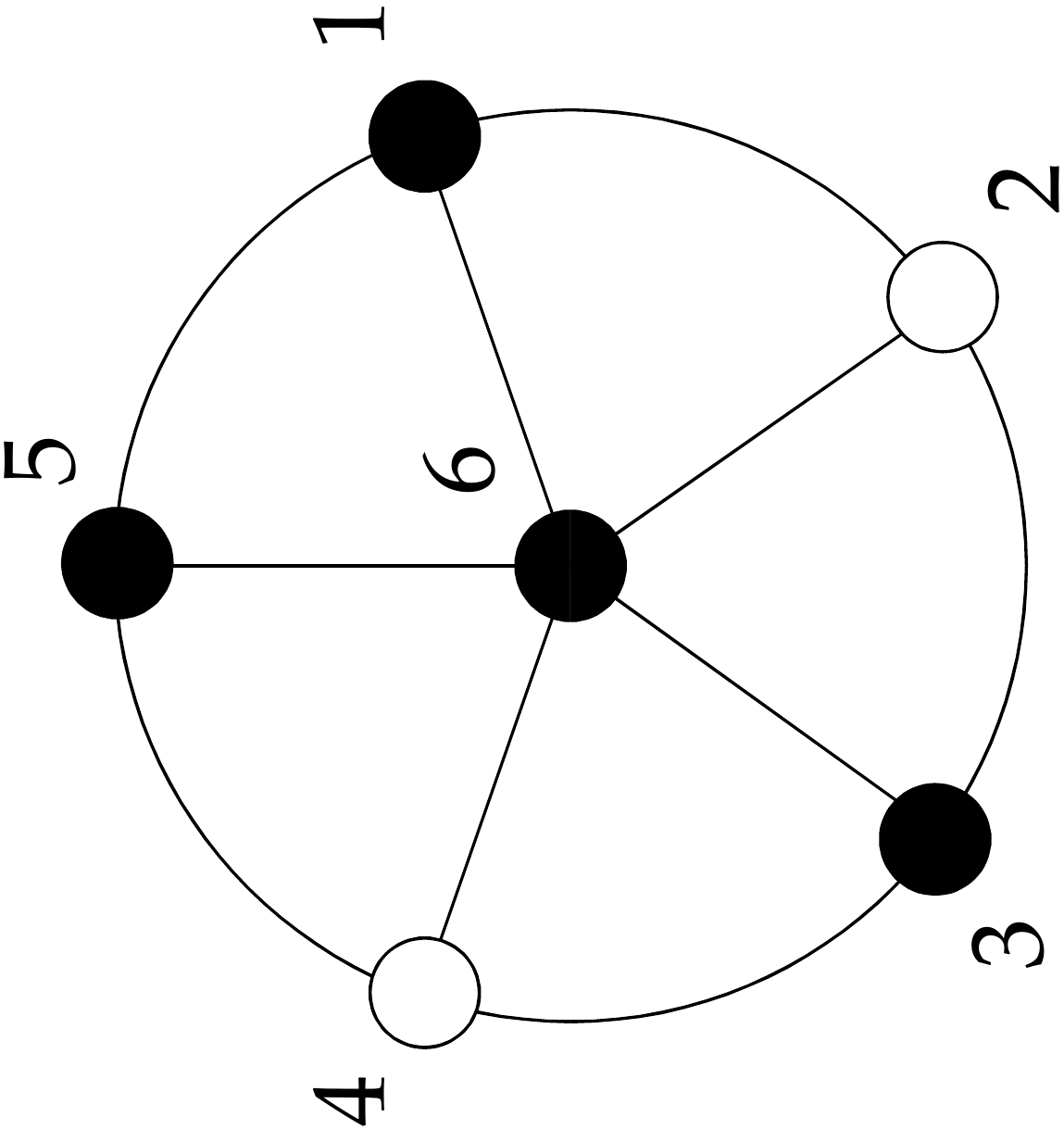}
\end{tabular}
\caption{\label{s5w6}Left: Star graph $S_{5}$. Star graphs $S_{N}$ are equivalent to complete bipartite graphs $K_{1,N}$. Right: Wheel graph $W_{6}$.}
\end{figure}

The investigation of interacting particle systems on graphs is an active field of mathematical reasearch \cite{woess,diaconis,saloff,caputo}. In the physics literature, however, exact results for exclusion processes on general graphs remain scarce. We showed that, besides on the usual one-dimensional chains under open and periodic boundary conditions, also known, respectively, as the path and cycle graphs $P_{N}$ and $C_{N}$, the SSEP($G$) is also amenable to investigation on other types of graphs with familiar techniques like quantum angular momentum algebra and basic group representation theory. The list of graphs that can be explored in this way includes star graphs---the star graph $S_{N}$ is just the complete bipartite graph $K_{1,N}$---, wheel graphs, and finite regular trees; see figure~\ref{s5w6}.

We avoided employing SU($N$) representations and Young tableaux on purpose to keep the exposition elementary. Indeed, the study of permutation invariant operators like (\ref{weiss}) and (\ref{nqweiss}) is more natural by means of the permutation group; see, e.\,g., \cite{hamer} for background and \cite{salerno,popkov,boyd} for closely related applications. In the totally asymmetric exclusion process (TASEP) on the one-dimensional periodic chain, the degeneracies of the eigenspectrum were also investigated directly from the Bethe ansatz equations and many combinatorial formul\ae\ were found relating the number of multiplets and their degeneracies with the size $N$ of the chain and number $n$ of particles in the system \cite{golinelli}. Some of the results found there can be explained by the invariance of the process under the action of the ``spin flip'' operator (\ref{flip}) in the half-filled case together with reflection ($i \to N-i$) and permutation symmetries. Although exactly solvable, the infinitesimal generator of the TASEP, which in quantum spin language corresponds to an XXZ model with an imaginary Dzyaloshinskii-Moriya term, precludes a straightforward application of the SU($2$) machinery as it was developed here to the clarification of its eigenspectrum structure; in this case, a full-fledged analysis in terms of the permutation group becomes necessary.

Finally, it is clear that interacting spin waves, variational states (including matrix product ans\"{a}tze), and cluster approximations, among other approaches, could also be applied in the investigation of exclusion processes on graphs, e.\,g., to estimate the gap of the mixed-spins operator (\ref{mixed}). An investigation of the stationary particle density and current fluctuations of the SSEP($G$) in general is also of some interest and we intend to resume this subject soon.

\section*{Acknowledgements}

The author thanks Prof.\ Ant\^{o}nio R. Moura (UFU, Brazil) for helpful conversations, Yeva Gevorgyan (YSU, Armenia) for support and friendship, and CNPq, Brazil, for partial financial support under grant PDS 151999/2010-4.


\section*{Appendix: Outer multiplicities in the Clebsch-Gordan series for \linebreak arbitrary spin-$s$ representations of the SU($2$)}

The degeneracies of the eigenvalues of the operators (\ref{weiss}) and (\ref{nqweiss}) partly come from the existence of many possible different combinations of the elementary spins summing up to a definite value of $S$. This degeneracy is encoded in the outer multiplicity $d_{s}(N,S)$ of the irreducible representation $\mathcal{D}^{(S)}$ appearing in the $N$-fold tensor product
\begin{equation}
\label{tensor}
[\mathcal{D}^{(s)}]^{\otimes N} = 
\bigoplus_{S=S_{\rm min}}^{sN} d_{s}(N,S) \mathcal{D}^{(S)}
\end{equation}
resolved into the direct sum via repeated application of the Clebsch-Gordan series \cite{hamer}
\begin{equation}
\label{cbseries}
\mathcal{D}^{(\ell)} \otimes \mathcal{D}^{(\ell')} = \mathcal{D}^{(|{\ell'-\ell}|)} \oplus \mathcal{D}^{(|{\ell'-\ell}|+1)} \oplus \cdots  \oplus \mathcal{D}^{(\ell'+\ell)},
\end{equation}
where $S_{\rm min}=1/2$ if $s$ is half-integer and $N$ is odd and $S_{\rm min}=0$ otherwise, $N \geq 2$ is the number of vertices of the graph, and $s$ is the magnitude of the spins involved.

To calculate $d_{s}(N,S)$, we first notice that in a system of $N$ spins $s$, the number $b_{s}(N,M)$ of states of total magnetization $M$ is given by the coefficient of $z^{M}$ in the expansion of $(z^{-s} + z^{-s+1} + \cdots + z^{s})^{N}$. This recipe stems from the solution of the simple combinatorial problem of distributing $M$ things among $N$ boxes each supporting a minimum of $-s$ and a maximum of $+s$ things \cite{vilenkin}. Next we notice that $d_{s}(N,S)$ and the numbers $b_{s}(N,M)$ are related by
\begin{equation}
\label{dbb}
d_{s}(N,S) = b_{s}(N,S) - b_{s}(N,S+1),
\end{equation}
since in a given subspace of fixed $M$ we have $S \geq |{M}|$, so that $b_{s}(N,S) - b_{s}(N,S+1)$ counts just those states with exactly total spin $S$. Finally, an explicit expression for $b_{s}(N,S)$ can be obtained from its generating function,
\begin{equation}
\label{cnm}
\begin{split}
(z^{-s}+z^{-s+1}+\cdots+z^{s})^{N} &= \sum_{M=-sN}^{+sN}b_{s}(N,M)z^{M} = \\
= z^{-sN}(1+z+\cdots+z^{2s})^{N} &= z^{-sN}\sum_{M=0}^{2sN}c_{2s+1}(N,M)z^{M},
\end{split}
\end{equation}
such that $b_{s}(N,M) = c_{2s+1}(N,sN+M)$. It is clear from (\ref{cnm}) that $b_{s}(N,M)=b_{s}(N,-M)$. The coefficients $c_{2s+1}(\cdot,\cdot)$, that are a variant of the usual multinomial coefficients, are known as generalized or extended binomial coefficients of order $2s+1$, and reduce to the standard binomial coeficients when $s=\frac{1}{2}$ \cite{vilenkin,bollinger},
\begin{equation}
\label{binomial}
b_{1/2}(N,M) = c_{2}(N,{\textstyle\frac{1}{2}}N+M) = {N \choose \frac{1}{2}N+M};
\end{equation}
compare (\ref{dbb}) and (\ref{binomial}) with (\ref{deg}). If we put $z=1$ in (\ref{cnm}) we obtain $\sum_{M} b_{s}(N,M) = (2s+1)^{N}$, as required.

It turns out that generalized binomial coefficients can be written in terms of standard binomial coeficients \cite{vilenkin,bollinger}. The resultant expression for $b_{s}(N,M)$ is
\begin{equation}
\label{bin}
b_{s}(N,M) = 
\sum_{k \geq 0} (-1)^{k} {N \choose k} {(s+1)N+M-(2s+1)k-1 \choose sN+M-(2s+1)k},
\end{equation}
where the summation runs over $k$ as long as the summing terms are non-null. Both the upper and the lower terms in the second binomial coefficient above are integer, even if $N$ is odd and $s$ is half-integer, because then $M$ will necessarily be half-integer. Equations (\ref{dbb}), (\ref{cnm}), and (\ref{bin}) together solve the Clebsch-Gordan series decomposition (\ref{tensor}) for arbitrary spin-$s$ representations of the SU($2$).



\begin{thebibliography}{99}
\bibitem{note}An example of ferromagnetic mixed-spins chain is the MnNi(NO$_2$)$_4$\,(ethylenediamine)$_2$ compound, that displays an alternating arrangement of ferromagnetically coupled Mn$^{\rm II}$ ($S=5/2$) and Ni$^{\rm II}$ ($S=1$) ions \cite{kahn}. Even in this case, weak antiferromagnetic interactions between neighboring chains drive the bulk material to an antiferromagnetic stack of ferromagnetic sheets below $T_{\rm Neel} \simeq 2.45$\,K.
\end{thebibliography}
\end{document}